\documentclass[fleqn,usenatbib]{mnras}
\usepackage{newtxtext,newtxmath}

\usepackage[T1]{fontenc}

\DeclareRobustCommand{\VAN}[3]{#2}
\let\VANthebibliography\thebibliography
\def\thebibliography{\DeclareRobustCommand{\VAN}[3]{##3}\VANthebibliography}

\usepackage{mnras_macros}
\usepackage{graphicx}	
\usepackage{amsmath}	

\title[Evolution of Magorrian relation in $\nu^2$GC]{The evolution of supermassive blackhole mass--bulge mass relation by a semi-analytic model, $\nu^2$GC}

\author[Shimizu et al.]{Tatsuki Shimizu,$^{1}$\thanks{shimizu@astro1.sci.hokudai.ac.jp}
Taira Oogi,$^{2,3}$
Takashi Okamoto,$^{4}$\thanks{takashi.okamoto@sci.hokudai.ac.jp}
Masahiro Nagashima,$^{5}$
Motohiro Enoki$^{6}$
\newauthor
\\
$^{1}$Department of Cosmosciences, Graduate School of science, Hokkaido University, N10 W8, Kitaku, Sapporo, 060-0810, Japan\\
$^{2}$Department of Electrical and Computer Engineering, National Institute of Technology, Asahikawa College, Shunkodai 2-2-1-6, Asahikawa, 071-8142, Japan\\
$^{3}$Research Center for Space and Cosmic Evolution, Ehime University, Bunkyo-cho, Matsuyama, Ehime, 790-8577, Japan\\
$^{4}$Faculty of Science, Hokkaido University, N10 W8, Kitaku, Sapporo, 060-0810, Japan\\
$^{5}$Faculty of Education, Bunkyo University, 3337, Minami-ogishima, Koshigaya, Saitama, 343-8511, Japan\\
$^{6}$Center for General Education, Tokyo Keizai University, 1-7-34, Minami-cho, Kokubunji, Tokyo, 185-8502, Japan\\
}

\date{Accepted XXX. Received YYY; in original form ZZZ}

\pubyear{2022}

\begin{document}
\label{firstpage}
\pagerange{\pageref{firstpage}--\pageref{lastpage}}
\maketitle

\begin{abstract}
We have investigated the redshift evolution of the relationship between supermassive black hole (SMBH) mass and host bulge mass using a semi-analytical galaxy formation model $\nu^2$GC.
Our model reproduces the relation in the local universe well. 
We find that, at high redshift ($z \gtrsim 3$), two sequences appear in the SMBH mass--bulge mass plane.  
The emergence of these two sequences can be attributed to the primary triggers of the growth of the SMBHs and bulges: galaxy mergers and disc instabilities. The growth of SMBHs and bulges as a result of galaxy mergers is responsible for giving rise to the high-mass sequence, in which SMBHs are more massive for a given host bulge mass than in the low-mas sequence.  Conversely, disc instabilities are accountable for the emergence of the low-mass sequence.
At lower redshifts, galaxy mergers tend to become increasingly deficient in gas, resulting in a preferential increase of bulge mass without a corresponding growth in SMBH mass. This has the effect of causing galaxies in the upper sequence to shift towards the lower one on the SMBH mass-bulge mass plane. The galaxies that undergo dry mergers serve to bridge the gap between the two sequences, eventually leading to convergence into a single relation known in the local universe. Our results suggest that the observations of the SMBH mass-bulge mass relation in high redshifts can provide insight into their growth mechanisms.
\end{abstract}




\section{Introduction}
It is generally accepted that most galaxies house a supermassive black hole (SMBH) at their centres. Studies have demonstrated that the mass of the SMBH is correlated with physical properties of its host galaxy's bulge such as stellar mass and velocity dispersion \citep{magorrian98, ferrarese00, haring04, McConnell13}, giving rise to the concept of co-evolution between SMBHs and their host bulges. These correlations between the SMBH mass and host bulge properties can be utilized to infer the formation processes of the SMBHs. For instance, \citet{shirakata16} proposed that the majority of seed black holes must possess a mass as small as $10^3~M_\odot$ in order to account for the relationship between SMBH mass and its host bulge mass in local dwarf galaxies.


In this study, we focus on the evolution with redshift of the relationship between the mass of an SMBH and the mass of its host bulge, in order to discern the type of constraints that can be placed on the co-evolution mechanism through a comparison between theoretical predictions and observations. Observationally, it is known that these quantities are proportional in the local universe, referred to as the $M_\mathrm{BH}$--$M_\mathrm{bulge}$ relation. However, recent observations of high-redshift quasars have revealed a different relationship, with luminous quasars having SMBHs that are overmassive for a given bulge mass when compared to the local relation. 
For example, \citet{ding20} found that the ratio of SMBH mass to total stellar mass (bulge mass) is $\sim$2.7 ($\sim$7) times larger at $z\sim1.5$ than in the local Universe.
However, there are studies that report no evolution in the $M_\mathrm{BH}$--$M_\mathrm{star}$ relation compared to the local relation in the range $0 <~ z <~ 2.5$ (e.g. \citealt{cisternas2011}; \citealt{schramm2013}; \citealt{suh2020}), although the former two studies find an evolution in the $M_\mathrm{BH}$--$M_\mathrm{bulge}$ relation.
Recently, \citet{YZhang23} have obtained the $M_\mathrm{BH}$--$M_\mathrm{star}$ relation relation at $z\sim2$, extending low SMBH masses of $M_\mathrm{BH} \sim 10^7 - 10^8 M_{\odot}$, and have suggested the same trend at $z\sim1$, i.e. overmassive SMBHs for a given $M_\mathrm{star}$ compared to the local relation.
At higher redshiftrs of $z\sim6$, luminous quasars with overmassive SHBMs have also been observed \citep{wang16, decarli18, shimasaku19}, whereas low-luminosity quasars have SMBHs that are consistent with or less massive than the local relation \citep{willott17, izumi18, izumi19, izumi21}.

There have been several theoretical studies conducted in the field of SMBH growth. One such study was proposed by \citet{wyithe03}, which suggested a model of SMBH growth that is self-regulated by active galactic nucleus (AGN) feedback. The authors predicted a redshift-dependent $M_\mathrm{BH}$--$M_\mathrm{bulge}$ relation, where $M_\mathrm{BH}/M_\mathrm{bulge}$ $\propto$ $[\xi(z)]^{1/2}(1 + z)^{3/2}$, with $\xi(z)$ being a weak function of redshift.
Another study by \citet{croton06b} utilized a semi-analytic model of galaxy formation to investigate the evolution of the $M_\mathrm{BH}$--$M_\mathrm{bulge}$ relation. He demonstrated that the relationship evolves with redshift, with a higher black hole mass associated with a given bulge mass at higher redshifts, as compared to the present day. He assumed that galaxy mergers are the primary drivers for both bulge and SMBH growth. 
While the studies described above focused on galaxy mergers, our previous studies have shown that disc instabilities play a significant role in causing the bulge and SMBH growth at high redshift, especially in faint AGNs \citep{nu2GCagn, oogi2020}. To fully understand the evolution of the $M_\mathrm{BH}$--$M_\mathrm{bulge}$ relation, we need to address both processes and study their roles in the evolution of the relation.
\citet{zhang23} proposed an empirical model that infers the statistical connection between dark matter haloes, galaxies, and supermassive black holes, and used the model to study the evolution of the $M_\mathrm{BH}$--$M_\mathrm{bulge}$ relation. Their model predicted that the normalisation and the slope of the $M_\mathrm{BH}$--$M_\mathrm{bulge}$ relation increases mildly from $z=0$ to 10.

Cosmological hydrodynamic simulations also predict different evolutions of the $M_\mathrm{BH}$--$M_\mathrm{bulge}$ relation (or $M_\mathrm{BH}$--$M_\mathrm{star}$, where $M_\mathrm{star}$ is the total stellar mass in the galaxy), depending on the subgrid models employed. 
{\citet{habouzit2021, habouzit2022} investigated the redshift evolution of the median $M_\mathrm{BH}$--$M_\mathrm{star}$ relation using six large-scale cosmological hydrodynamic simulations. Their study reveals different trends among these simulations: the normalisation of the relation shows a decreasing trend with decreasing redshift in Illustris \citep{vogelsberger_2014_Illustris}, Horizon-AGN \citep{dubois_2014_horizon_agn}, and EAGLE \citep{schaye_2015_eagle}. In contrast, TNG100, TNG300 \citep{weinberger_2017_IllustrisTNG, pillepich_2018_IllustrisTNG} and SIMBA \citep{dave_2019_simba} show an increasing trend. }
In \citet{forever22}, SMBHs grow slowly at $M_\mathrm{star} \lesssim 10^{10}~M_\odot$, and then their mass steeply increases as the galaxies become more massive. A similar trend can be seen in \citet{rosas-guevara16}. On the other hand, SMBH masses in galaxies $M_\mathrm{star} \lesssim 10^{10}~M_\odot$ obey the local $M_\mathrm{BH}$--$M_\mathrm{star}$ relation even at high redshifts in \citet{weinberger16}.
As an alternative approach, \citet{inayoshi2022} discuss the evolutionary path of a massive seed BH through the region above the local relation using radiation hydrodynamical simulations resolving nuclear scales (see also \citealt{hu2022} for the evolution to overmassive SMBHs at $z\sim6$).

We aim to provide a theoretical prediction of the $M_\mathrm{BH}$--$M_\mathrm{bulge}$ relation at high redshifts. Understanding how this relationship evolves over time will allow us to place constraints on the formation and evolution of SMBHs. To achieve this, we use the $\nu^2$GC semi-analytic model, which incorporates results from $N$-body simulations to model the formation and evolution histories of dark matter haloes and uses phenomenological equations to model the complicated baryon physics \citep{nu2GC, nu2GCagn}.

In this model, we mainly consider two mechanisms as the growth process of bulges and SMBHs \citep{nu2GCagn}: galaxy mergers and disc instabilities. We demonstrate how these two growth mechanisms are reflected in the redshift evolution of the $M_\mathrm{BH}$--$M_\mathrm{bulge}$ relationship.

This paper is structured as follows: First, we briefly explain the $\nu^2$GC model, particularly the models for the evolution of the bulge and SMBH, in Section 2. Subsequently, in Section 3, we show the predictions by our model and delve into the causes for the presence of the two sequences in the $M_\mathrm{BH}$--$M_\mathrm{bulge}$ relations at high-redshifts, and how they merge into a single sequence towards low-redshift. Finally, we summarize our results in Section 4. 

\section{MODEL DESCRIPTIONS}

In this section, we provide a brief overview of the semi-analytic model employed, $\nu^2$GC. The formation histories of dark matter haloes (merger-trees) are extracted from $N$-body simulations, while phenomenological equations are employed to model baryonic physics, including gas cooling, star formation, supernova feedback, SMBH growth, and AGN feedback. 
{This model has been extensively tested against a variety of observational data. \citet{nu2GCagn} demonstrated its ability to explain stellar mass functions of galaxies at $z < 4$, AGN luminosity functions at $z < 4$, the local SMBH mass function, and the local $M_\mathrm{BH}$--$M_\mathrm{bulge}$ relation. We also examined the Eddington ratio distribution in \citet{shirakata_edd_dis} and found it to be consistent with the available data. In addition, we examined and confirmed its agreement with the Soltan argument in \citet{shirakata_soltan}. \citet{oogi2016} delved into quasar clustering and found qualitative agreement with observational results. Subsequently, using more extensive $N$-body data, \citet{oogi2023} explored AGN luminosity functions and the cosmic variance.}
In this section, for baryon physics, the description is restricted to the physics relevant to the evolution of bulges and SMBHs. The full details of the model can be found in \citet{nu2GC} and \citet{nu2GCagn}.
Throughout this paper we use the default parameter set of \citet{nu2GCagn}.

\subsection{Merger-trees}
The merger-trees used in this paper are constructed from a cosmological $N$-body simulation known as "micro-Uchuu" \citep{uchuu21}. The specifications of this simulation are as follows: The total number of particles, $N$, is $640^3$, the box length, $L$, is 100 $h^{-1}$ Mpc, the softening length, $\epsilon$, is 4.27 $h^{-1}$ kpc, and a mass of a dark matter particle, $m_p$, is $3.27\times10^8$ $h^{-1}~M_\odot$.

\subsection{Growth of bulges and SMBHs}

We assume that a bulge increases its mass by violent relaxation of the stellar components due to mergers and disc instabilities, as well as by starbursts. Galaxy mergers and disc instabilities also trigger the starbursts. We also assume that a fraction of the cold gas is accreted by a central SMBH during the starburst. An SMBH also increases its mass by absorbing other SMBHs brought in by merging galaxies, although the contribution of this process to the mass growth is small. 
Below we describe in detail how galaxy mergers and disc instabilities lead to mass increases for both bulges and SMBHs.

\subsubsection{Galaxy mergers}

Galaxies can merge through two mechanisms in our model: dynamical friction and random collisions. Dynamical friction leads to a merger between a central galaxy and a satellite galaxy, while random collisions can result in a merger between two satellite galaxies.
 We estimate the timescale for both processes using the same method as \citet{nu2GC}. 
The more massive galaxy is considered the primary galaxy having the mass\footnote{In this paper we define the mass of a galaxy as the sum of stellar mass, cold gas mass, and SMBH mass.}, $M_1$, and the less massive is the secondary galaxy with the mass of $M_2$. 
After a merger, stars and cold gas from the secondary galaxy are added to the bulge of the primary galaxy. At the same time, the primary galaxy's bulge also acquires stars and cold gas from its own disc. The amount of stars obtained from the disc is given by $\Delta M_\mathrm{1ds} = \min(f_\ast$$M_2, M_\mathrm{1ds})$, where $f_\ast = G(\mu) = 2\mu/(1 + \mu)$ with $\mu = M_2/M_1$ \citep{hopkins09a}.

The gas within a radius of $R_\mathrm{gas}$ loses its angular momentum efficiently and dissipates into a central starburst. To determine $R_\mathrm{gas}$, we use Eq.~(7) from \cite{hopkins09a}, which takes the following form:
\begin{equation}
\frac{R_\mathrm{gas}}{r_\mathrm{ds}} = (1 - f_\mathrm{1g})f_\mathrm{1d}F(\theta, b)G(\mu),
\label{eq:r_gas}
\end{equation}
where $r_\mathrm{ds}$ is the scale radius of the disc, assumed to be the same for the gas and stellar discs. The mass fraction $f_\mathrm{1g}$ represents the fraction of gas mass in the primary disc, and $f_\mathrm{1d}$ represents the fraction of the primary galaxy's disc. The function $F(\theta, b)$ depends on the peri-centric distance $b$ before the merger and the inclination of the secondary galaxy's orbit relative to the primary galaxy's disc, $\theta$. 
Since we cannot obtain $b$ and $\theta$ from merger trees of the dark matter haloes, we employ the average value of $F(\theta, b)$ suggested by \citet{hopkins09a}, $\langle F(\theta, b)\rangle = 1.2$. 
Assuming that the disc has an exponential surface density profile, the cold gas mass within   $R_\mathrm{gas}$, which is supplied to the bulge and is exhausted by a starburst, is 
\begin{equation}
\Delta M_{1, \mathrm{dg}} = M_{1, \mathrm{dg}}\left\{1 - \left(1 + \frac{R_\mathrm{gas}}{r_\mathrm{ds}}\right)\exp\left(-\frac{R_\mathrm{gas}}{r_\mathrm{ds}}\right)\right\}, 
\label{eq:delta_m_burst}
\end{equation}
where $M_{1,\mathrm{dg}}$ is the cold gas mass in the primary's disc. 
According to Eq.~(\ref{eq:r_gas}) and (\ref{eq:delta_m_burst}), $\Delta M_{1, \mathrm{dg}}$ is always less than $M_{1, \mathrm{dg}}$. 
This indicates that the formation of pure bulge galaxies is not possible via galaxy mergers.
To allow the formation of a pure bulge galaxy as a major merger remant, we assume that the disc of the primary is completely destroyed when $\mu > \mu_\mathrm{major} = 0.89$ as in \citet{nu2GCagn}. 

Due to the merger-induced starburst, the bulge experiences an increase in its stellar mass of $\Delta M_\mathrm{star, burst}$, which is determined by the following equation:
\begin{equation}
{\Delta}M_\mathrm{star,burst} = \frac{\alpha}{\alpha + \beta + f_\mathrm{BH}}M^{0}_\mathrm{cold},
\label{eq:dMc_merger}
\end{equation}
where $M^{0}_\mathrm{cold}$ denotes the mass of cold gas present in the bulge just after the merger, which can be expressed as:
\begin{equation}
M^0_\mathrm{cold} \equiv \Delta M_{1, \mathrm{dg}} + \sum_{i > 1} M_{i, \mathrm{dg}}.
\end{equation}
The parameter $\alpha$ represents the locked-up mass fraction, while $\beta$ indicates the strength of supernova feedback and is defined as $\beta \equiv (V_\mathrm{d}/V_\mathrm{hot})^{-\alpha_\mathrm{hot}}$, where $V_\mathrm{d}$ is the disc rotation velocity. In this paper, we use $V_\mathrm{hot} = 121.64~\rm{km~s}^{-1}$and $\alpha_\mathrm{hot} = 3.62$. The fraction of gas that is accreted by the SMBH is $f_\mathrm{BH} = 0.02$.

\subsubsection{disc instability}

When a galactic disc becomes gravitationally unstable, a part of the disc undergoes violent relaxation and is added to the bulge. 
According to \citet{eln82}, a galactic disc becomes gravitationally unstable when the following condition is satisfied
\begin{equation}
\frac{V_\mathrm{max}}{(GM_\mathrm{disc}/r_\mathrm{ds})^{\frac{1}{2}}} < \epsilon_\mathrm{DI,crit},
\end{equation}
where $V_\mathrm{max}$ is the maximum rotational velocity, $r_\mathrm{ds}$ is the scale length and $r_\mathrm{ds} = (1/\sqrt{2})\langle {\lambda}_\mathrm{H} \rangle R_\mathrm{init}$ where $R_\mathrm{init}$ is the initial radius of the hot gas sphere and $\langle {\lambda}_\mathrm{H}\rangle$ is the mean value of the dimensionless spin parameter. We use $\langle {\lambda}_\mathrm{H} \rangle = 0.042$ \citep{bett07}. 
We use $\epsilon_\mathrm{DI,crit}$ = 0.75 to reproduce the observed SFR density of the universe.

The fraction of the stellar disc mass that is added to the bulge is determined by a parameter, $f_\mathrm{bar}$, and we assume $f_\mathrm{bar} = 0.63$. 
The mass of the cold gas that loses its angular momentum and dissipates into a central starburst due to the disc instability is determined by the equation similar to Eq.~(\ref{eq:dMc_merger}) as 
\begin{equation}
{\Delta}M_\mathrm{dg}^\mathrm{DI} = M_\mathrm{dg}\left\{1 - \left(1 + \frac{R_\mathrm{gas}}{r_\mathrm{ds}}\right)\exp\left(-\frac{R_\mathrm{gas}}{r_\mathrm{ds}}\right)\right\},
\label{eq:dMburst_DI}
\end{equation}
where 
\begin{equation}
\frac{R_\mathrm{gas}}{r_\mathrm{ds}} = (1 - f_\mathrm{g})f_\mathrm{d}f_\mathrm{bar}.  
\label{eq:dMc_DI}
\end{equation}
Here, we omit the subscript 1 since only one galaxy is involved in this process. The variables have the same meaning as in Eqs.~(\ref{eq:r_gas}) and (\ref{eq:delta_m_burst}).

\subsubsection{SMBH growth}

Seed BHs with a mass of $10^{3}M_\odot$ are promptly embedded in the newly formed galaxy.
During a starburst triggered by a galaxy merger or disc instability, the supermassive black hole (SMBH) gains the following mass increment:
\begin{equation}
{\Delta}M_\mathrm{BH} = \frac{f_\mathrm{BH}}{\alpha + \beta + f_\mathrm{BH}}M_\mathrm{cold}^{0}.
\end{equation}
The SMBH in the primary galaxy also increases its mass by absorbing other SMBHs brought in by merged satellite galaxies.

\section{Result}
\begin{figure}
	\includegraphics[width=\columnwidth]{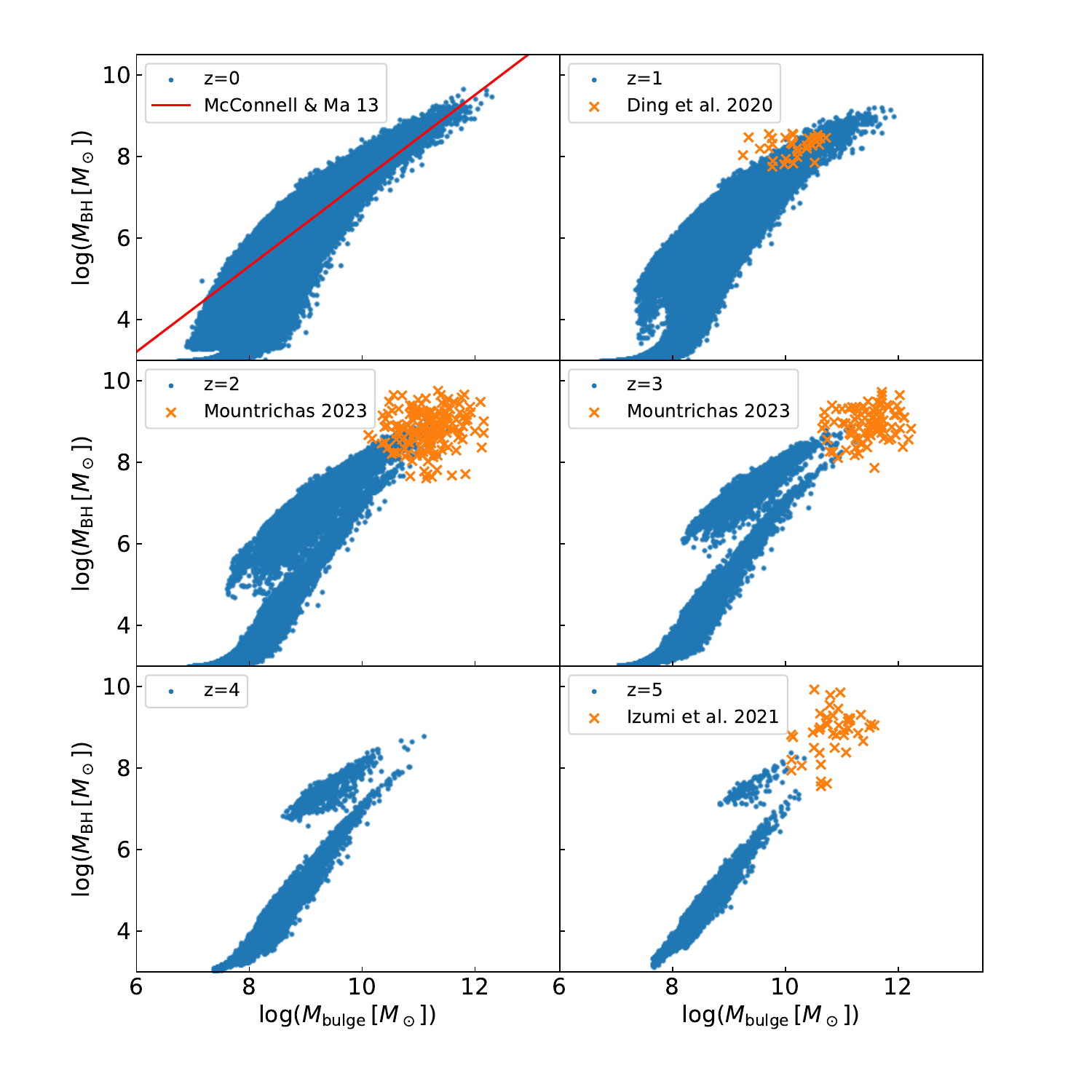}
    \caption{The correlation between SMBH masses and their host bulge masses across the redshift range of $z=0$ to 5, with the aid of the model prediction. The solid line in the top-left panel indicates the best-fitting power law for the local galaxies by \citet{McConnell13}. The orange crosses represent observational estimates within specific redshift ranges. For redshifts between $1$ and $2$, the estimates are from \citet{ding20} (top-right panel). The middle-left panel displays estimates for redshifts between $1.5$ and $2.2$ from \citet{Mountrichas2020}, while the middle-right panel shows estimates for redshifts greater than $2.2$ from the same source. The bottom-right panel represents estimates for redshifts greater than $6$ from \citet{izumi21}. It is important to note that \citet{Mountrichas2020} measured the total stellar mass, while \citet{izumi21} estimated the dynamical mass of the host galaxies instead of the bulge mass.
    }
    \label{fig:Mbh-Mbulge}
\end{figure}
%

Fig.~\ref{fig:Mbh-Mbulge} shows the model predictions for the relationship between SMBH mass and host bulge mass in the redshift range $0$ to $5$. For $z = 0$, our model reproduces the local relationship well. 
Our local $M_\mathrm{BH}$--$M_\mathrm{bulge}$ is curved. This bend is mainly due to the stellar feedback as discussed in \citet{shirakata16} (see also \citet{fontanot2015}).
We find that the growth of the bulges and SMBHs via disc instabilities also contributes to making the relation curved, as we demonstrate in Appendix~\ref{appendix-a}.
Our model matches the observed estimates of SMBHs and galaxies at high redshifts. 
Our model fails to explain the presence of very massive SMBHs and their associated bulges at high redshifts, which may be a consequence of the limited size of our simulation box.
Our model prediction shows a remarkable characteristic: the presence of two sequences emerges prominently at high redshifts ($z \gtrsim 3$), while only a single sequence persists at $z = 0$.
This result differs significantly from the prediction of \citet{croton06b}, which suggests that the slope of the $M_\mathrm{BH}$--$M_\mathrm{bulge}$ relationship remains almost constant, with only the normalisation increasing at higher redshifts. A similar trend to \citet{croton06b} is reported by \citet{enoki14}. 
These models consider only galaxy mergers as a triggering mechanism for bulge formation and gas accretion onto SMBHs. 
Comparatively, the results of our model suggest that disc instabilities also play an important role in the early evolution of SMBHs \citep{nu2GCagn}. Therefore, we expect that the presence of two mechanisms, galaxy mergers and disc instabilities, for bulge and SMBH growth is the reason behind the existence of the two sequences.
In what follows, we will explore the origin of these two sequences and examine how they gradually converge to establish the $M_\mathrm{BH}$--$M_\mathrm{bulge}$ relation in the local universe.

\begin{figure}
	\includegraphics[width=\columnwidth]{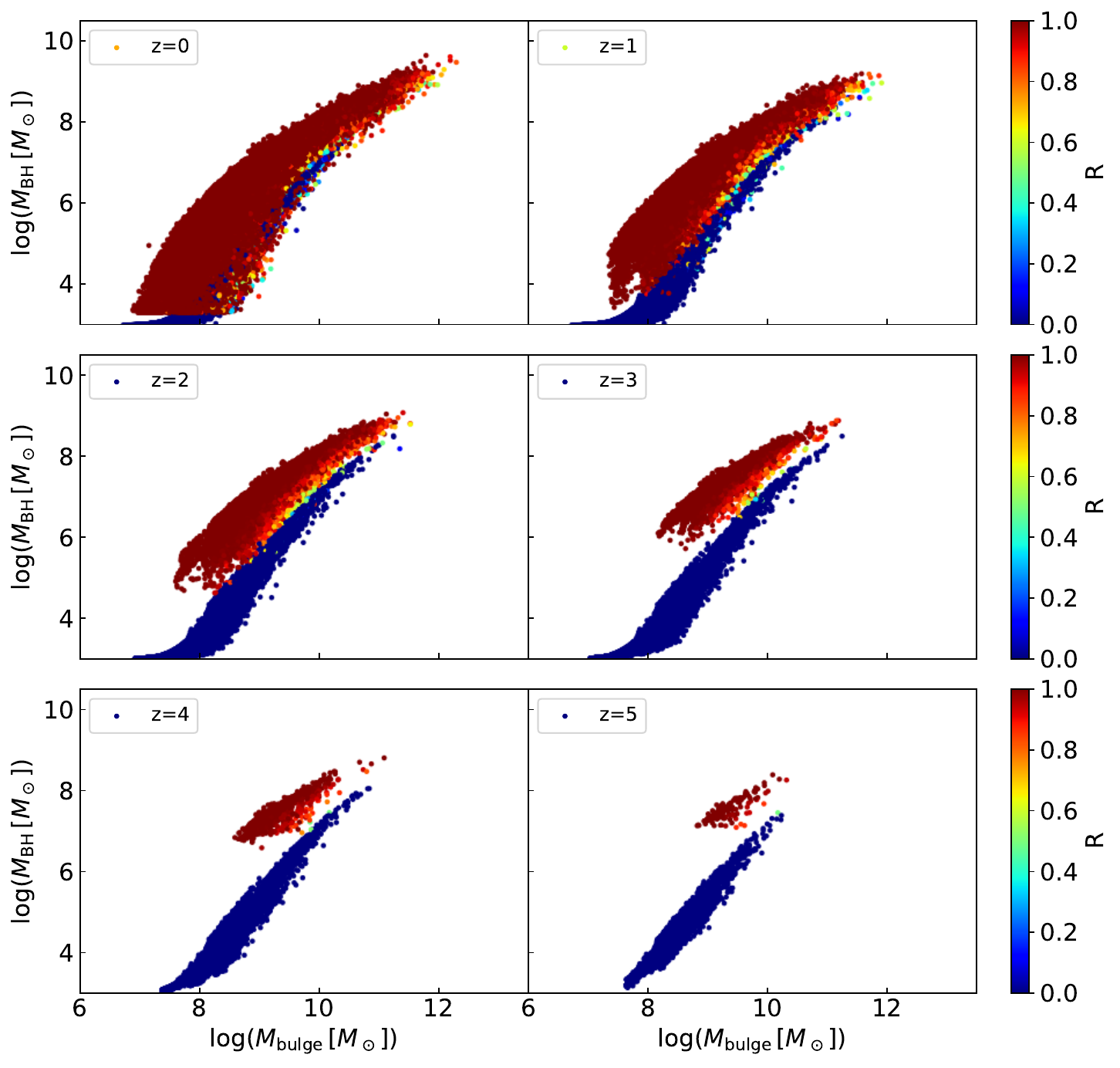}
    \caption{The same as Fig.~\ref{fig:Mbh-Mbulge} but the points are colour-coded based on the ratio of mass gained via merger-induced accretion to the sum of the merger-induced and disc instability-induced accreted mass: the merger-induced mass fraction, $\Delta M_\mathrm{mrg}/(\Delta M_\mathrm{mrg} + \Delta M_\mathrm{DI}$). 
    }
    \label{fig:Mbh merger Mbh di_figure}
\end{figure}

To investigate the relative importance of two processes, we calculated the mass of an SMBH gained through gas accretion triggered by galaxy mergers, denoted as $\Delta M_\mathrm{mrg}$, and that obtained through disc instabilities, denoted as $\Delta M_\mathrm{DI}$. Fig.~\ref{fig:Mbh merger Mbh di_figure} presents the SMBHs colour-coded based on the ratio of mass gained through merger-induced accretion to the total accreted mass induced by mergers and disc instabilities, as expressed by:
\begin{equation}
R=\frac{\Delta M_\mathrm{mrg}}{\Delta M_\mathrm{mrg} + \Delta M_\mathrm{DI}}.
\end{equation}
This ratio represents the relative contribution of gas accretion triggered by galaxy mergers versus disc instabilities.
This figure shows that most of the SMBHs in the upper sequence have acquired their mass through accretion triggered by galaxy mergers. In contrast, the SMBHs in the lower sequence have acquired their mass through accretion triggered by disc instabilities. We can infer that the reason for the existence of the two sequences is due to these different accretion mechanisms. 
It is also apparent from this figure that the lower sequence does not exhibit significant redshift evolution, whereas the upper sequence broadens as redshift decreases, eventually bridging the gap between the two sequences.

\begin{figure} 
\includegraphics[width=\columnwidth]{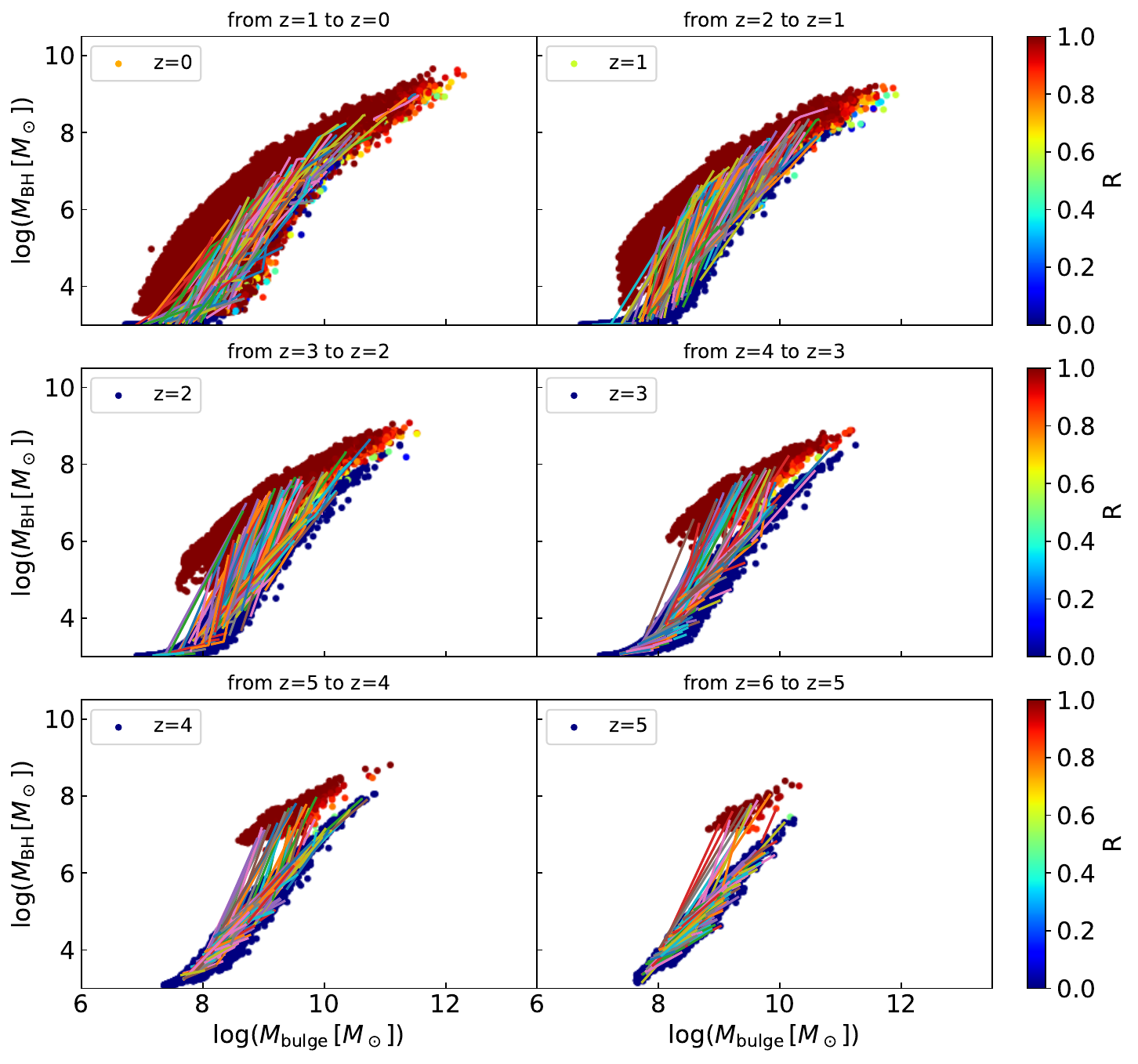}
    \caption{Evolution of galaxies in the lower sequence in the $M_\mathrm{BH}$--$M_\mathrm{bulge}$ plane. 
    We randomly select galaxies from the lower sequence at redshift $z+1$ and connect their positions at redshift $z$ using the solid lines.
    The redshift $z$ is indicated by the figure legend. 
    }
    \label{fig:evolution of lower galaxies_figure}
\end{figure}

To demonstrate the evolution of individual galaxies (or SMBHs) in the $M_\mathrm{BH}$--$M_\mathrm{bulge}$ plane, we randomly select galaxies from the lower sequence at redshift $z+1$ and identify their positions in the plane at redshift $z$. Fig.~\ref{fig:evolution of lower galaxies_figure} shows each evolutionary path from $z+1$ to $z$. We observe that while some galaxies in the lower sequence evolve along this sequence, others jump to the upper sequence. This jump is attributed to merger-driven gas accretion, as suggested by the merger-induced fraction.
On the other hand, the evolution along the lower sequence is caused by the disc instabilities. 
The shallow slope of evolutionary paths caused by disc instabilities suggests that the ratio of mass acquired by an SMBH to that acquired by its host bulge through disc instabilities is smaller than the ratio due to mergers.
This implies that, in the disc instability, the mass of the disc stars that are added to the bulge is much larger than the mass of the cold gas that is fueled to the central starburst. 

\begin{figure}
	\includegraphics[width=\columnwidth]{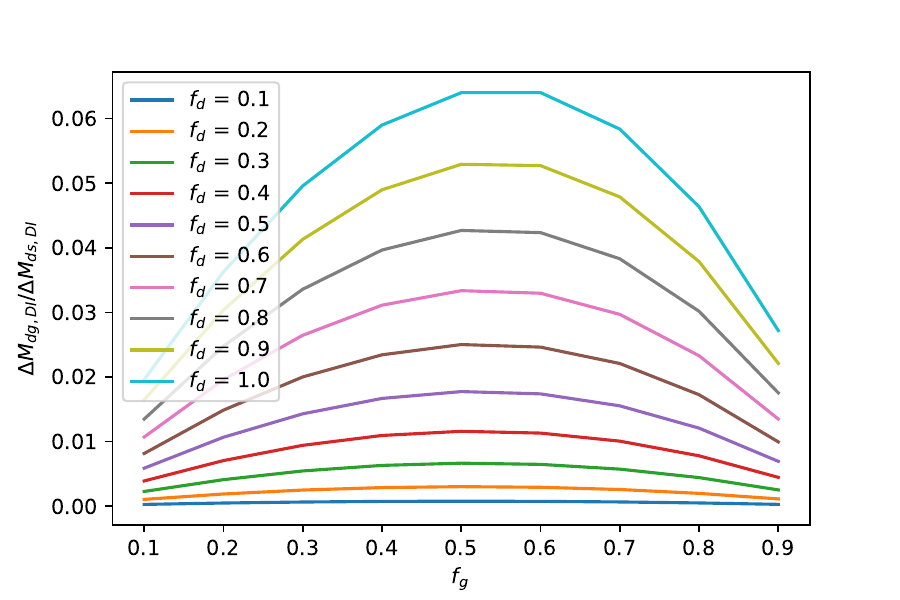}
    \caption{
    The ratio between the mass of stars that migrate from a disc to a bulge, $\Delta M_\mathrm{ds}^\mathrm{DI}$, and the mass of gas consumed by a starburst, $\Delta M_\mathrm{dg}^\mathrm{DI}$, in the case of disc instability as a function of the gas mass fraction in the disc. 
    We show the ratio by changing the disc mass fraction, $f_\mathrm{d}$, from 0.1 to 1.0. 
    }
    \label{fig:ratio of gas from disc to stars from disc_figure}
\end{figure}

Equation (\ref{eq:dMburst_DI}) can be used to compute the ratio between the mass of stars that migrate from a disc to a bulge, $\Delta M_\mathrm{ds}^\mathrm{DI}$, and the mass of gas consumed by a starburst, $\Delta M_\mathrm{dg}^\mathrm{DI}$, in the case of disc instability. In Fig.~\ref{fig:ratio of gas from disc to stars from disc_figure}, we present $\Delta M_\mathrm{dg}^\mathrm{DI}/\Delta M_\mathrm{ds}^\mathrm{DI}$ as a function of the gas fraction, $f_\mathrm{g}$, for various values of disc mass fraction, $f_\mathrm{d}$.
From this figure, it is apparent that $\Delta M_\mathrm{dg}^\mathrm{DI}/\Delta M_\mathrm{ds}^\mathrm{DI}$ consistently remains small (less than 0.1) . 
While we use a similar equation for the merger-induced gas supply (Eq.~(\ref{eq:delta_m_burst})), the ratio of stellar mass added to the bulge and gas consumed by the central starburst can be much larger than in the case of disc instability. This is because all the gas in the secondary galaxy is utilized for the starburst. Additionally, in the case of a major merger with $\mu > 0.89$, all the gas in the merging system is available for the starburst. 
Therefore galaxies tend to jump up to the upper sequence through mergers.

\begin{figure}
	\includegraphics[width=\columnwidth]{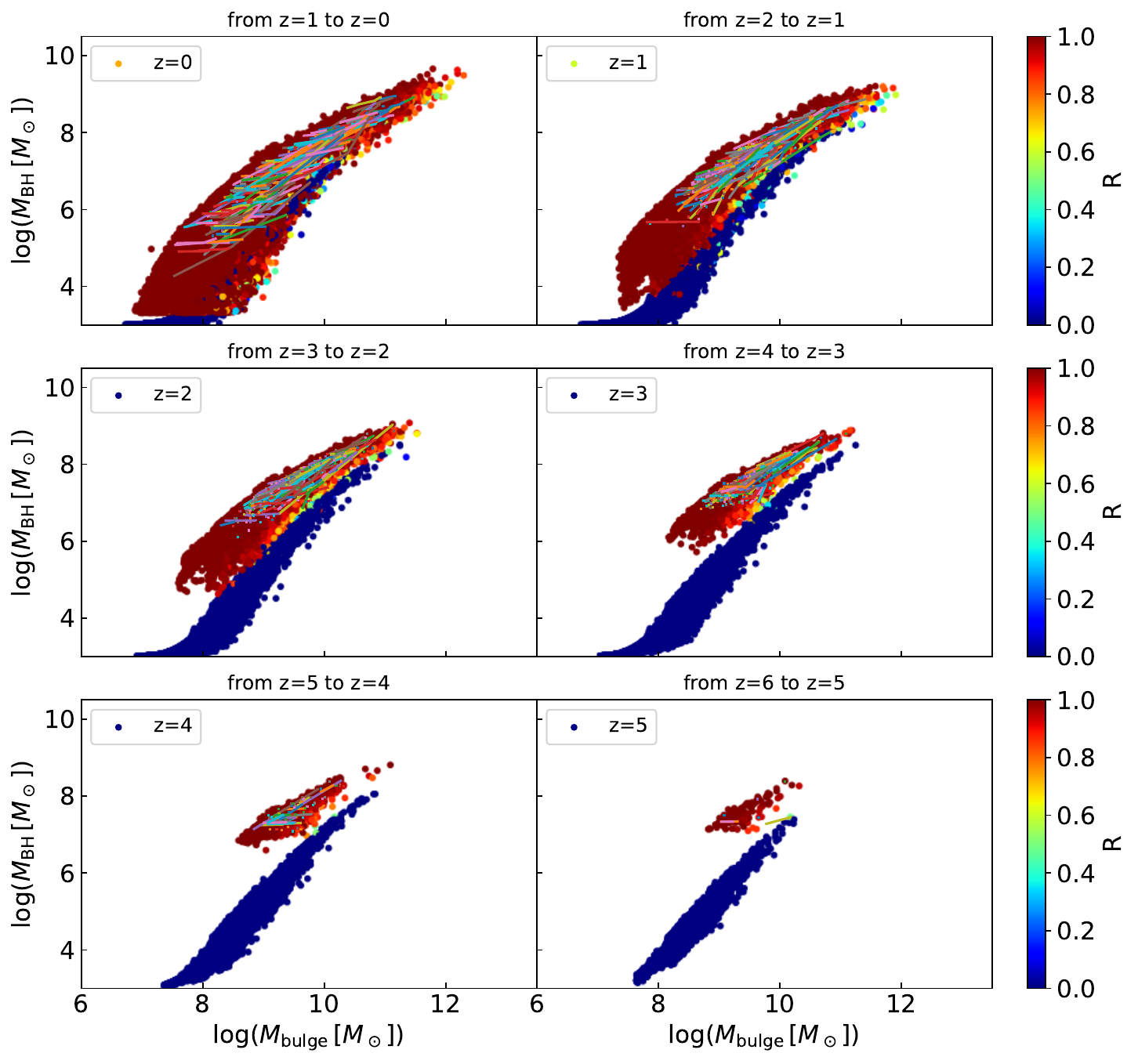}
    \caption{
    Evolution of galaxies in the upper sequence in the $M_\mathrm{BH}$--$M_\mathrm{bulge}$ plane. Galaxies at reshift $z +1$ are randomly selected and connected to their position in the plane at $z$ by solid lines. 
    }
    \label{fig:evolution of upper galaxies_figure}
\end{figure}

From Fig.~\ref{fig:evolution of lower galaxies_figure}, we also observe that galaxies in the lower sequence at redshift $z + 1$ tend to either stay in the lower sequence or move up to the upper sequence at redshift $z$, without filling the gap.
Hence, galaxies initially in the upper sequence must be the ones that fill the gap.
We perform the same analysis as in Fig.\ref{fig:evolution of lower galaxies_figure} on galaxies from the upper sequence at redshift $z+1$. Specifically, we randomly select galaxies from the upper sequence at $z+1$ and identified their positions in the $M_\mathrm{BH}$--$M_\mathrm{bulge}$ plane at redshift $z$. The resulting evolutionary paths are shown in Fig.\ref{fig:evolution of upper galaxies_figure}.

We find that galaxies located in the upper sequence at redshift $z+1$ do not transition to the lower sequence at redshift $z$; instead, they remain in the upper sequence. 
However, some galaxies exhibit trajectories with slopes shallower than those of the upper sequence itself. These galaxies broaden the upper sequence, and over time, the two sequences eventually merge into a single sequence. Our findings are consistent with previous semi-analytic studies \citep[e.g.,][]{croton06b, enoki14}, which suggest that at lower redshifts, a lower SMBH mass is associated with a given bulge mass. 
As galaxies become progressively gas-poor towards lower redshifts, an SMBH cannot increase its mass much, while a bulge can grow by acquiring stars in discs. Consequently, the evolution of galaxies in the $M_\mathrm{BH}$--$M_\mathrm{bulge}$ plane becomes nearly horizontal.

\section{Summary and discussion}

\begin{figure}
	\includegraphics[width=\columnwidth]{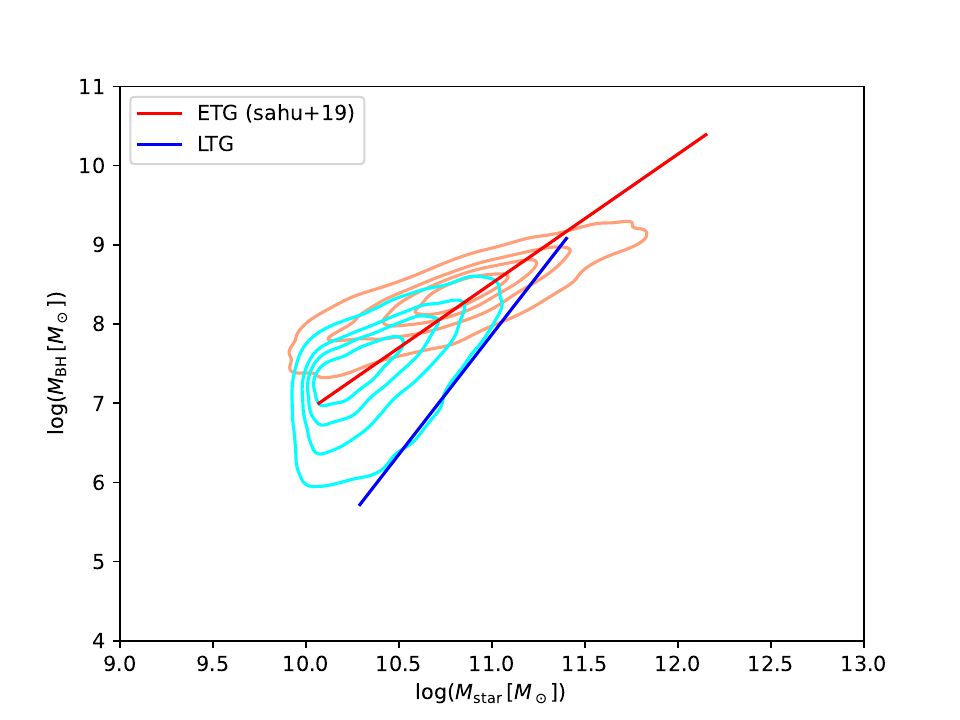}
    \caption{
    {$M_\mathrm{BH}$--$M_\mathrm{star}$ relations for early-type galaxies (ETGs) and late-type galaxies (LTGs). The red and blue contours show our model prediction for ETGs ($B/T \ge 0.5$) and LTGs ($B/T < 0.5$), respectively. The red and blue solid lines show the linear fits for the observed ETGs and LTGs from \citet{sahu2019}. }
    }
    \label{fig:mol_fig}
\end{figure}

We have investigated how the relation between the mass of SMBHs and their host bulges, commonly known as the $M_\mathrm{BH}$--$M_\mathrm{bulge}$ relation, changes with redshift. We have used a semi-analytic model called $\nu^2$GC for this purpose. We find that the $M_\mathrm{BH}$--$M_\mathrm{bulge}$ relation exhibits two distinct sequences at high redshifts ($z \gtrsim 3$), while reproducing the local relation well.
We identify the origins of these two sequences: the sequence in which SMBH mass is greater for a given bulge mass is established by galaxy mergers that induce both bulge and SMBH growth, whereas SMBHs and bulges that primarily gain their mass through disc instabilities are located in the lower SMBH mass sequence.

A dry merger occurs when two galaxies merge without significant gas content, causing an increase in bulge mass by adding stars from disrupted discs to the bulge. However, the SMBH mass does not increase much due to the limited availability of gas. As a result, when a galaxy in the upper SMBH mass sequence undergoes a dry merger, it tends to move towards the gap between the two sequences. As galaxies become increasingly depleted of gas towards lower redshifts, the effect of dry mergers becomes more pronounced. Eventually, the two sequences merge to form a single local relation.
This redshift evolution of the upper sequence is consistent with the evolution of the $M_\mathrm{BH}$--$M_\mathrm{bulge}$ relation primarily driven by galaxy mergers \citep{croton06b, enoki14}.

According to our model, there is no clear single relationship between the SMBH mass and the bulge mass with respect to the whole BH population at high redshifts. At $z \lesssim 1$, a familiar single correlation appears. 
Our model also suggests that a fraction of undermassive SMBHs at high redshifts do not remain in the region below the local relation, but have experienced merger-driven growth to become overmassive SMBHs and have moved to the region above the local relation. Fig.~\ref{fig:Mbh merger Mbh di_figure} shows that SMBHs that have grown purely through disc instabilities dominate the region just above the lower envelope of the $M_\mathrm{BH}$--$M_\mathrm{bulge}$ relation at $z\sim 1-2$. The host bulge of these SMBHs may have some characteristic properties in terms of kinematics or morphology. While studies of the morphology of AGN host galaxies have begun at $z\sim 1$ (\citealt{ding20}), these host quantities are difficult to observe in current studies at $\sim 2$ (\citealt{YZhang23}) due to limited spatial and spectral resolution and the difficulty of AGN-host galaxy decomposition. However, {\it James Webb Space Telescope (JWST)} will advance studies of the physical properties of AGN host galaxies. They can be used to validate our model with detailed observations of the host galaxies. We will investigate them in future studies.

{While our model predicts a single sequence of the $M_\mathrm{BH}$--$M_\mathrm{bulge}$ relation at $z = 0$, the two sequences at high redshifts may leave some traces in the relation at z=0. For example, \cite{Menci2023} showed that their model predicts a higher Eddington ratio for a lower mass BH for a given stellar mass. However, we find no such correlation in the local $M_\mathrm{BH}$--$M_\mathrm{bulge}$ relation predicted by our model \citep[see also][]{oogi_2017}. Observationally, \citet{sahu2019} suggested that early- and late-type galaxies obey different scaling relations.  Since it is possible that galaxies having merger-driven bulges and those having DI-driven bulges have different morphologies, we plot the $M_\mathrm{BH}$--$M_\mathrm{star}$ relations of the bulge-dominated and disc-dominated galaxies in Fig.~\ref{fig:mol_fig}, where we classify galaxies with a bulge-to-total mass ratio (B/T) greater than 0.5 as bulge-dominated galaxies and the rest as disc-dominated galaxies. We find that our disc-dominated galaxies obey a steeper $M_\mathrm{BH}$--$M_\mathrm{star}$ relation than the bulge-dominated galaxies. This result is in qualitative agreement with the finding of \citet{sahu2019}, and therefore the steeper relation observed for late-type galaxies may suggest that the main gas supply mechanism to the galactic centre is different for the early- and late-type galaxies. A more quantitative comparison with the data will require further refinement of our model to obtain the Hubble types of the model galaxies. We have also confirmed that our result does not change qualitatively when we change the B/T threshold.}

The $M_\mathrm{BH}$--$M_\mathrm{bulge}$ relation splitting into two sequences at high redshifts is a result of our model of bulge growth and gas accretion onto an SMBH triggered by disc instability. However, the process of how disc instabilities transform a (part of a) disc into a spheroid and how much gas is funnelled to the central starburst is not yet fully understood. This means that different modelling approaches may yield different results from ours. For instance, \citet{lacey16}  assume that the whole disc is turned into a spheroid, and all the gas in the galaxy is available for the central starburst, with a fraction of it accreting to the SMBH. Since this treatment is almost identical to the case of a major merger, we do not expect their model to produce two distinct sequences in the $M_\mathrm{BH}$--$M_\mathrm{bulge}$ relation.

While semi-analytic studies that focus on the redshift evolution of $M_\mathrm{BH}$--$M_\mathrm{bulge}$ are limited, there are a lot of predictions for the local relation from other semi-analytic models \citep[e.g.][]{Hirschmann2012, Menci2014, fontanot2015, Fontanot2020}. Since the local relation is used to constrain these models, their predictions are generally similar to our own. 
Our model predicts a break in the $M_\mathrm{BH}$-$M_\mathrm{bulge}$ relation, similar to what has been reported in previous studies \citep{fontanot2015, Fontanot2020}. This break appears to be related to stellar feedback, as discussed in Appendix~\ref{appendix-a} and by earlier authors \citep{fontanot2015, cirasuolo2005}. However, in our model, the break occurs at a lower bulge mass compared to \citet{fontanot2015} and \citet{Fontanot2020}. This difference likely arises from the different treatments of stellar feedback in bulges between the models.
\citet{Menci2014} compared $M_\mathrm{BH}$-$M_\mathrm{bulge}$ relations from models with different black hole accretion mechanisms (galaxy interactions vs. disc instabilities). They found that the relation is lower and steeper in the disc instability model compared with the interaction model, consistent with our high-redshift results.
In contrast, the relations from \citet{guo2011} and \citet{Hirschmann2012} did not show such a break, despite having similar treatments of SMBH growth, stellar feedback, and AGN feedback. This suggests that small differences in model details can lead to variations in the shape of the $M_\mathrm{BH}$--$M_\mathrm{bulge}$ relation. As an example, we show that modifying the disc instability model can yield less curved relations in Appendix~\ref{appendix-a}. More insights may come from studying the redshift evolution of this scaling relation in different models, as we have done in this paper.

{
Predictions of the $M_\mathrm{BH}$--$M_\mathrm{star}$ relation from cosmological hydrodynamic simulations are particularly interesting since such simulations inherently include all the physical processes responsible for the gas supply to galactic centres and the violent relaxation of disc stars. \citet{habouzit2021} studied the redshift evolution of the $M_\mathrm{BH}$--$M_\mathrm{star}$ relation using six cosmological hydrodynamic simulations (Illustris, TNG100, TNG300, Horizon-AGN, EAGLE, and SIMBA).
In Illustris, Horizon-AGN and EAGLE, SMBHs are on average overmassive at high redshifts compared to local relations. In contrast, in TNG100, TNG300 and SIMBA, SMBHs are undermassive at high redshifts. This discrepancy highlights the critical influence of subgrid models within the simulations and shows the lack of theoretical consensus regarding the redshift evolution of the scaling relation between SMBH mass and galaxy stellar mass. Interestingly, EAGLE and Horizon-AGN show a subtle hint of the existence of a lower sequence at high redshifts, while we expect that these simulations need higher resolution to correctly capture the disc instabilities and their effects on galaxy evolution. 
}

Given that the redshift evolution of the $M_\mathrm{BH}$--$M_\mathrm{bulge}$ relation relies on assumptions regarding bulge formation and gas accretion onto an SMBH, there is potential for high redshift observations of the bulge and SMBH masses to constrain these physical processes.
Although, the number of such observations is increasing \citep[e.g.][]{wang16, willott17, decarli18, shimasaku19, izumi18, izumi19, izumi21}, it is still too early to confirm or refute the model. 
Dedicated simulations to study how disc instability drives bulge formation and gas inflow to the galactic centre quantitatively will also help to improve the model.


\section*{Acknowledgements}
{We appreciate the detailed review and useful suggestions by the anonymous referee, which have improved our paper.}
This study is supported by JSPS/MEXT KAKENHI Grants (21H04496, 20H05861, 19H01931, 20H01950, 18H05437, 23K03460, 21H05449, and 20K22360). 
We thank Instituto de Astrofisica de Andalucia (IAA-CSIC), Centro de Supercomputacion de Galicia (CESGA) and the Spanish academic and research network (RedIRIS) in Spain for hosting Uchuu DR1 and DR2 in the Skies \& Universes site for cosmological simulations. The Uchuu simulations were carried out on Aterui II supercomputer at Center for Computational Astrophysics, CfCA, of National Astronomical Observatory of Japan, and the K computer at the RIKEN Advanced Institute for Computational Science. The Uchuu DR1 and DR2 effort has made use of the skun@IAA\_RedIRIS and skun6@IAA computer facilities managed by the IAA-CSIC in Spain (MICINN EU-Feder grant EQC2018-004366-P).


\section*{Data Availability}
The data underlying this article can be shared upon reasonable request to the corresponding author.








\appendix

\section{The dependence of $M_\mathrm{BH}$--$M_\mathrm{bulge}$ relation at $z=0$ on some parameters}
\label{appendix-a}

\begin{figure}
	\includegraphics[width=\columnwidth]{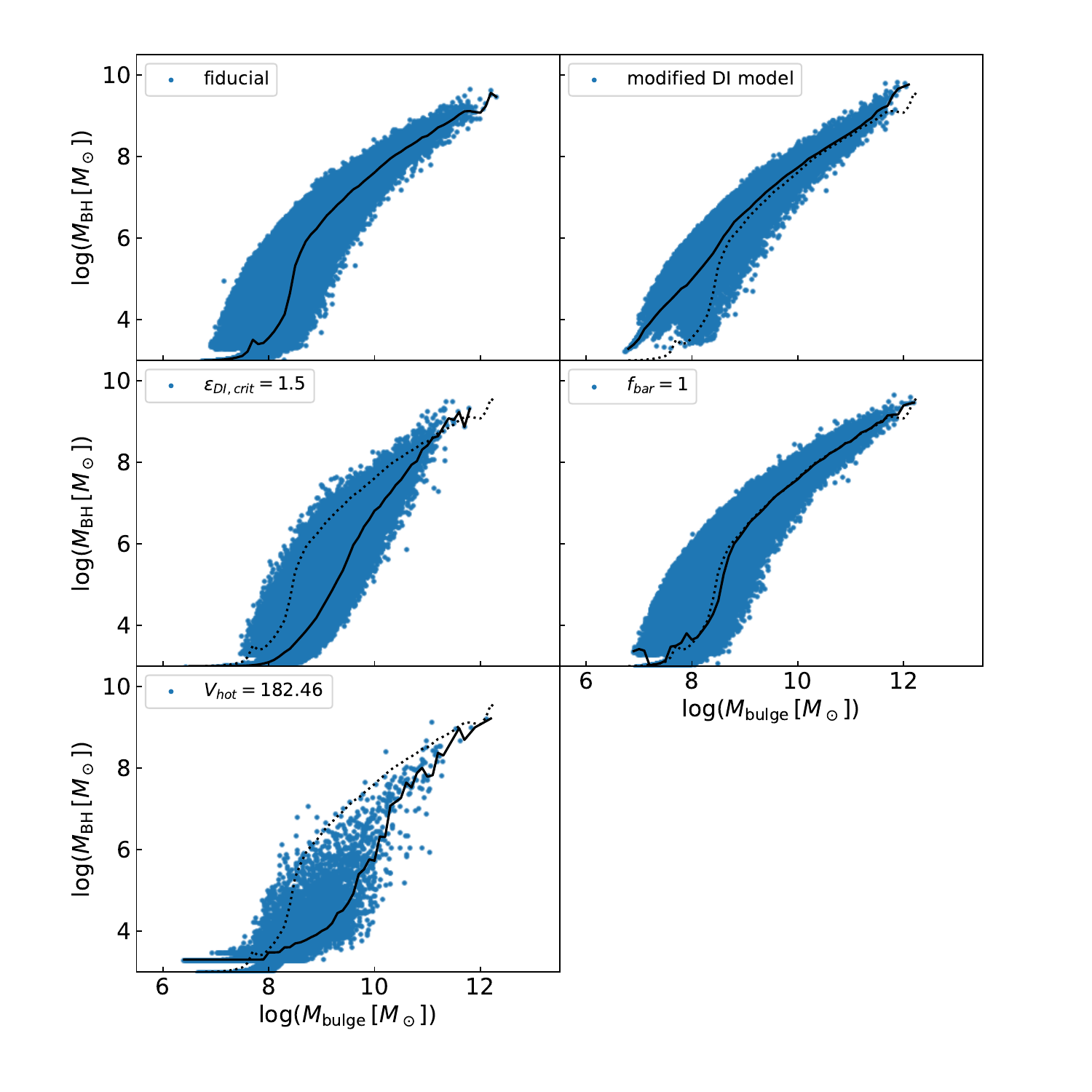}
    \caption{
    The local $M_\mathrm{BH}$--$M_\mathrm{bulge}$ relations for some variants of our model. The top-left panel shows the fiducial model for reference.
The top-right panel displays a model in which the gas mass supplied to the galactic centre during disc instability is increased (see the text for details).
In the middle-left panel, we present a model where the frequency of disc instabilities is increased by doubling the value of $\epsilon_\mathrm{DI,crit}$. The middle-right panel shows a model with $f_\mathrm{bar} = 1$, i.e., the fraction of disc mass that migrates to the bulge during disc instability is increased.
The bottom-left panel presents a model in which the value of $V_\mathrm{hot}$ is 1.5 times larger than in the fiducial model, corresponding to stronger stellar feedback. In each panel, the solid line indicates the median relation and the dotted line indicates the relation for the fiducial model.}
    \label{fig:change}
\end{figure}

To investigate the impact of adopted parameter values and models related to the growth of SMBHs and bulges on the local $M_\mathrm{BH}$--$M_\mathrm{bulge}$ relation at $z=0$, we examine how this relation changes when several of them are modified. It seems that our curved relation is originated by the stellar feedback and disc instability-driven SMBH and bulge growth, we thus focus on these two processes. Since our model only considers the radio mode as AGN feedback, varying the AGN feedback efficiency primarily affects the number and maximum masses of massive galaxies and SMBHs. Therefore, we do not modify AGN feedback parameters here. 
The top-right panel of Fig.~\ref{fig:change} shows the $M_\mathrm{BH}$--$M_\mathrm{bulge}$ relation at $z=0$ for a model where $\Delta M_\mathrm{dg}^\mathrm{DI} = \min(M_\mathrm{dg}, \Delta M_\mathrm{ds}^\mathrm{DI})$ instead of Eq.~\ref{eq:dMburst_DI} (the gas fraction transferred from a disc to a bulge during disc instability is significantly higher than the fiducial model).
Due to the more efficient SMBH growth via disc instabilities,  the relation becomes less curved compared with the fiducial model. The middle-left panel shows the relationship in a model where the frequency of disc instability is increased by doubling the value of $\epsilon_\mathrm{DI,crit}$. Increasing the importance of the disc instabilities makes the relation less curved and steeper than the fiducial model. The middle-right panel shows the relation by a model with $f_\mathrm{bar} = 1$. Changing this parameter has almost no effect on the shape of the local relation as discussed in \citet{shirakata16}. The lower-left panel shows the relation by a model where $V_\mathrm{hot}$ is increased by a factor of 1.5 (i.e. stronger stellar feedback).  Due to stronger stellar feedback, the bent occurs at a higher bulge mass. 

\bsp	
\label{lastpage}
\end{document}